\documentclass[preprint,showpacs,showkeys,nobibnotes,nofootinbib,eqsecnum,floats,unsortedaddress,byrevtex,aps,epsfig]{revtex4}
\usepackage{amssymb}
\usepackage{amsfonts}

\input{tcilatex}
\input epsf

\begin{document}

\title{SUM RULES FOR FOUR-SPINON DYNAMIC STRUCTURE FACTOR IN XXX MODEL}
\author{B. Si Lakhal}
\email{silakhal@wissal.dz}
\affiliation{D\'{e}partement de Physique, Universit\'{e} de Blida, BP 270 Blida 09000,
Algeria}
\author{A. Abada}
\email{a.abada@uaeu.ac.ae}
\affiliation{Physics Department, Faculty of Science, United Arab Emirates University, POB
17551 Al Ain, United Arab Emirates}
\altaffiliation{On leave from: D\'{e}part.~de Physique, Ecole Normale Sup\'{e}rieure, BP 92
Vieux-Kouba, 16050 Alger, Algeria}
\date{\today }

\begin{abstract}
In the context of the antiferromagnetic spin 1/2 Heisenberg quantum spin
chain (XXX model), we estimate the contribution of the exact four-spinon
dynamic structure factor $S_{4}$ by calculating a number of sum rules the
total dynamic structure factor $S$ is known to satisfy exactly. These sum
rules are: the static susceptibility, the integrated intensity, the total
integrated intensity, the first frequency moment and the nearest-neighbor
correlation function. We find that the contribution of $S_{4}$ is between
1\% and 2.5\%, depending on the sum rule, whereas the contribution of the
exact two-spinon dynamic structure factor $S_{2}$\ is between 70\% and 75\%.
This is consistent with the expected scattering weight of states from
outside the spin-wave continuum. The calculations are numerical and Monte
Carlo based. Good statistics are obtained.
\end{abstract}

\maketitle

\section{Introduction}

Quantum spin chains are in fact three-dimensional magnetic compounds in
which the magnetic interaction in one direction dominates over the\ two
others. They are not mere academic curiosities since many real-world
compounds do have this property. One such compound is KCuF$_{3}$, a
satisfactory realization of the one-dimensional spin 1/2 antiferromagnetic
Heisenberg model \cite{heisenberg}. Early description of its
crystallographic properties\ \cite{hirakawa-kurogi} confirms that it comes
in two types, (a) and (d) \cite{okazaki1,okazaki2}, both with similar spin
structures. The (antiferromagnetic) coupling constant along the chain
direction is roughly 100 times stronger than the (ferromagnetic) interchain
coupling constant, a feature confirmed by electron-spin resonance
measurements. This justifies amply a dominant one-dimensional behavior, and
additional specific heat measurements are consistent with this
interpretation. Also, an ideal strict one-dimensional magnetic chain will
not exhibit long-range order at any finite temperature $T$ \cite{baxter},
but realistic quasi-one-dimensional systems have a characteristic
temperature below which there is long-range order. For example, in KCuF$_{3}$
long-range order is manifest below 38K for type (a) and 20K for type (d) 
\cite{hirakawa-kurogi}.

Though presumably simpler than higher-dimensional systems, quantum spin
chains show strong and rich quantum behavior. If for example we consider
antiferromagnetic systems, we would classically anticipate a N\'{e}el state
traversed by spin-waves, and (linear) spin-wave theory is traditionally the
usual framework for such a description. In this context, the excitation
spectrum for the one-dimensional antiferromagnetic Heisenberg model is
predicted to be \cite{anderson1}:%
\begin{equation}
\omega _{\mathrm{cl}}\left( k\right) =2\left\vert \sin k\right\vert ,
\label{classical dispersion}
\end{equation}%
where $k$ is the momentum transfer along the chain direction and here, the
magnetic coupling constant is normalized to unity for convenience. But the
true ground state is actually different from the N\'{e}el state \cite{bethe}%
, and the lowest-lying excited states satisfy the so-called
des-Cloizeaux-Pearson (dCP) dispersion relation \cite{des cloizeaux-pearson}:%
\begin{equation}
\omega _{\mathrm{dCP}}\left( k\right) =\pi \left\vert \sin k\right\vert .
\label{dCP dispersion}
\end{equation}%
At first, these excitations were understood as spin-wave-like states with
spin one and early experiments on CPC \cite%
{endoh-shirane-birgeneau-richards-holt,heilmann-shirane-endoh-birgeneau-holt}
and KCuF$_{3}$ \cite%
{hutchings-ikeda-milne,satija-axe-shirane-yoshizawa-hirakawa} were
interpreted as a confirmation of this. But it was later shown \cite%
{faddeev-takhtajan,fowler} that the natural excitations of the model have
spin 1/2 and hence are fermions. These elementary excitations are called
spinons \cite{anderson2} and come always in pair. Furthermore, if the spin
of the system is an integer, the spinons are bound and lead to well-defined
spin-wave-like modes exhibiting a Haldane gap \cite{haldane}, a feature not
seen in the (linear) spin-wave theory. If the spin is a half-integer, the
spinons are unbound and there is no Haldane gap. Spin 1/2 compounds are even
more interesting in that the spectrum is not simply given by a definite
dispersion relation and indeed, analytic \cite{yamada} and finite chain \cite%
{bonner-sutherland-richards} calculations showed that it is actually a
continuum of excitations confined to first approximation, for a given $k$,
between a lower bound $\omega _{l}\left( k\right) $ and an upper bound $%
\omega _{u}\left( k\right) $ such that:%
\begin{equation}
\omega _{l}\left( k\right) =\omega _{\mathrm{dCP}}\left( k\right) ;\qquad
\omega _{u}\left( k\right) =2\pi \left\vert \sin \,k/2\right\vert .
\label{2 spinon continuum}
\end{equation}

The ground state properties and more particularly the excitation spectrum of
a magnetic system are analyzed using inelastic neutron scattering, the
cross-section of which is a function of the energy and momentum transfers $%
\omega $ and $\mathbf{k}$ respectively \cite{lovesey,squires}. All the above
mentioned features regarding the Heisenberg model are confirmed by
experiments. For example, inelastic neutron scattering on KCuF$_{3}$ shows
significant contribution to the scattering from regions consistent with the
spin-wave continuum (\ref{2 spinon continuum}) and not with linear spin-wave
theory, most particularly at high frequencies \cite%
{nagler-tennant-cowley-perring-satija}. This implies the inadequacy of
spin-wave theory \cite{kretzen-mikeska-patzak} at such high energies in the
quantum limit $s=1/2$, this even if the two-magnon interactions are taken
into account beyond the linear approximation \cite%
{holstein-primakoff,tennant-nagler-welz-shirane-yamada}. But there is
consistency at low-energies, and one must mention that spin-wave theory
works fine for compounds with high spins (classical limit) like KFeS$_{2}$
with $s=5/2$ where it gives accurate peak positions, line shapes and
relative intensities \cite{nagler-tennant-cowley-perring-satija}. One should
also note that inelastic neutron scattering on KCuF$_{3}$ showed consistency
of the field theory approach \cite{schulz}, valid only at long wavelengths,
particularly the temperature dependence of the scattering \cite%
{tennant-perring-cowley-nagler,tennant-cowley-nagler-tsvelik}. This is
important because it is an additional confirmation of the fermionic nature
of the elementary excitations, the spinons.

The inelastic neutron scattering is theoretically analyzed with the help of
the two-point dynamic structure factor (DSF) $S$, for, in the Born
approximation which is amply sufficient for the present purposes, the
differential cross-section per scattering solid angle $\Omega $ and outgoing
neutron energy $E_{f}$ is \cite{lovesey,squires}:%
\begin{equation}
\frac{d^{2}\sigma }{d\Omega \,dE_{f}}=N\,\sigma _{0}\frac{k_{f}}{k_{i}}%
\left\vert \frac{g}{2}F\left( \mathbf{k}\right) \right\vert
^{2}\sum_{i,j}\left( \delta _{ij}-\hat{k}_{i}\hat{k}_{j}\right)
\,S^{ij}\left( \omega ,\mathbf{k}\right) .  \label{cross section}
\end{equation}%
$N$ is the number of scatterers, $\sigma _{0}=0.2896$b a unit for magnetic
scattering, $g$ the Land\'{e} factor, $F\left( \mathbf{k}\right) $ the
magnetic form factor, $k_{f}$ the outgoing and $k_{i}$\ the incoming neutron
momenta respectively, $i$ and $j$ the cartesian coordinates. The DSF $S$ is
the Fourier transform of the two-point spin correlation function:%
\begin{equation}
S^{ij}\left( \omega ,\mathbf{k}\right) =\frac{1}{2\pi \hbar N}\int_{-\infty
}^{+\infty }dt\sum_{l,m}e^{i\left[ \mathbf{k\cdot }\left( \mathbf{r}_{l}-%
\mathbf{r}_{m}\right) -\omega t\right] }\,\left\langle S_{l}^{i}\left(
0\right) S_{m}^{j}\left( t\right) \right\rangle ,  \label{definition DSF}
\end{equation}%
where $\mathbf{r}_{l\left( m\right) }$ is the position of the spin $\mathbf{S%
}_{l\left( m\right) }$ on the chain. The Hamiltonian corresponding to the
antiferromagnetic isotropic Heisenberg model commutes with the total spin
operator, which implies that $S^{ij}\left( \omega ,\mathbf{k}\right) $ is
diagonal in $i$ and $j$ and $S^{ii}\left( \omega ,\mathbf{k}\right) $ is the
same for $i=x,y$ and $z$. The averaging in (\ref{definition DSF}) is
generally done at finite temperature, but at low temperatures, only the
ground state is retained. Also, for a quasi-one-dimensional system, the
cross-section for neutron scattering with momentum transfer $\mathbf{k}$
depends only on the component $k$ of $\mathbf{k}$ parallel to the chain.

Historically, the first attempts to calculate the DSF for the
antiferromagnetic Heisenberg model were made in the context of spin-wave
theory which, for the chain lying in the $z$-direction, gives the transverse
response as \cite{anderson1}:%
\begin{equation}
S_{\mathrm{SWT}}^{xx}\left( \omega ,k\right) =S_{\mathrm{SWT}}^{yy}\left(
\omega ,k\right) =\left\vert \tan \,k/2\right\vert \,\delta \left( \omega
-\omega _{\mathrm{cl}}\left( k\right) \right) ,  \label{SWT DSF}
\end{equation}%
and the longitudinal component $S_{\mathrm{SWT}}^{zz}\left( \omega ,k\right) 
$ with a logarithmic singularity at $\omega _{\mathrm{cl}}\left( k\right) $.
Non-isotropy here is a consequence of the assumption of a N\'{e}el ground
state with long-range order. Now the true quantum ground state does not have
long-range order, which means isotropy must apply. The inadequacy of (\ref%
{SWT DSF}) triggered further efforts. First those of \cite%
{muller-thomas-beck-bonner}, where, guided by exact results of the
one-dimensional XY model, numerical calculations on finite chains and known
sum rules, it is constructed an ansatz for the DSF of the spin $1/2$ chain
at zero temperature, the so-called M\"{u}ller ansatz:%
\begin{equation}
S_{\mathrm{Mul}}^{ii}\left( \omega ,k\right) =\frac{\Theta \left( \omega
-\omega _{l}\left( k\right) \right) \,\,\Theta \left( \omega _{u}\left(
k\right) -\omega \right) }{\sqrt{\omega ^{2}-\omega _{l}^{2}\left( k\right) }%
},  \label{Muller ansatz}
\end{equation}%
where $\Theta $ is the Heaviside step function. This form for the DSF has
two main features. (i) A square-root singularity at the lower boundary of
the spin-wave continuum. (ii) A built-in strict restriction to the spin-wave
continuum itself, which is not physical at the high-frequency end if the M%
\"{u}ller ansatz is to represent the total DSF. Indeed, the M\"{u}ller
ansatz predicts an abrupt high-energy cutoff while computer calculations 
\cite{muller-thomas-beck-bonner} suggest there is contribution of states
outside the continuum, particularly from above, but with a scattering weight
two orders of magnitude lower than that of neighboring states inside the
continuum. Within the spinon picture, the small contributions outside the
continuum are identified with processes in which more than two spinons are
created. There is also a relatively small but systematic underestimation of
the relative spectral weight near $k=\pi $ \cite%
{nagler-tennant-cowley-perring-satija}. But in overall, the M\"{u}ller
ansatz gives reasonable results (to order unity) for known sum rules, is
consistent with quantum Monte Carlo calculations \cite{deisz-jarrell-cox}
and supported by \cite{viswanath}. More importantly, it is in good agreement
with the inelastic neutron scattering experiments done on KCuF$_{3}$ \cite%
{nagler-tennant-cowley-perring-satija,tennant-perring-cowley-nagler,tennant-cowley-nagler-tsvelik}%
.

The second set of efforts was to map the antiferromagnetic Heisenberg model
in the long wavelength limit onto a relativistic quantum field theory and
exploit the bosonization method of Luther and Peschel \cite{luther-peschel}
to obtain an analytic expression of the DSF at finite temperature \cite%
{schulz}. For integer spins, one finds that the DSF has a single mode with a
gap as predicted in \cite{haldane}. For half-integer spins, the expression
of the DSF at zero temperature agrees with the low-momentum limit of the M%
\"{u}ller ansatz and\ at finite temperature, there is good agreement with
the measurements on KCuF$_{3}$ \cite%
{tennant-perring-cowley-nagler,tennant-cowley-nagler-tsvelik}.

But all the above results regarding the dynamic structure function, though
useful in their own right, are only approximate. This is important to note
because over the decades, a number of quantum spin chains, most notably the
Heisenberg model, have been amenable to exact solutions. As to the methods
used, first there was the Bethe ansatz \cite{bethe} which, as already
mentioned, gives the exact ground state for the model with characteristics
different from the classical N\'{e}el antiferromagnetic ordering. Then
techniques were developed in successful attempts to calculate exactly a
number of thermodynamic quantities, techniques like the method of
(commuting) transfer matrices and the Yang-Baxter equation \cite{baxter}.
These methods culminated in the so-called Quantum Inverse Scattering Method 
\cite{korepin-izergin-bogoliubov}, a milestone towards the recognition of
the quantum group symmetry present in the model. But up to here, all the
exact results were for the most part concerned with only the static
(thermodynamic) properties of the quantum spin chains; correlation functions
that encode the dynamics remained elusive to exact treatment. One had to
wait for advances in two-dimensional conformal field theory to see how the
infinite two-dimensional conformal symmetry allows the computation of
correlation functions of the so-called vertex operators, using bosonization
techniques \cite{difrancesco-mathieu-senechal}.\ The same strategy was then
applied to quantum spin chains once it became clear how to adapt the
bosonization method to deformed commutation relations between creation and
annihilation operators \cite%
{frenkel-jing,davies-foda-jimbo-miwa-nakayashiki,abada-bougourzi-gradechi,bougourzi1}%
. For the Heisenberg model, actual formal manipulations are made in the
context of the equivalent six-vertex model where vertex operators are
defined in the framework of the infinite-dimensional representation of the
quantum group. Correlation functions are then defined and, when mapped back
to local spin operators, compact expressions of form factors are obtained 
\cite{jimbo-miwa}.

The above treatment, though technically involved, opened the way for a
systematic exact treatment of the dynamic structure factor. In an expansion
in the (even) number of spinons \cite{bougourzi2}, it was first obtained an
exact expression for the two-spinon contribution $S_{2}$ \cite{BCK}, and a
comparison with the M\"{u}ller ansatz showed that it gives a better account
of the phenomenology \cite{KMB,BKM}. Next an exact expression of the
four-spinon contribution $S_{4}$ to the DSF was derived in \cite{ABS} and
its behavior as a function of the neutron energy and momentum transfers $%
\omega $ and $k$ respectively was studied in \cite{silakhal-abada}.

The present work aims at furthering the investigation of the properties of
the four-spinon DSF $S_{4}$. Through an estimation of a number of sum rules,
we address the issue of \ the spectral weight within $S_{4}$ in the total $S$%
. The sum rules we use are: the static susceptibility, the integrated
intensity, the total integrated intensity, the first frequency moment and
the nearest-neighbor correlation function. We reach the conclusion that $%
S_{4}$ contributes a weight between 1\% and 2.5\%, depending on the sum rule
used, whereas $S_{2}$ contributes a weight between 70\% and 75\%. This is
consistent with the finite chain calculations of \cite%
{muller-thomas-beck-bonner} where it is observed that states just up the
spin-wave continuum contribute with a scattering weight two orders of
magnitude smaller than states within the continuum. The sum rules we
investigate involve multi-dimensional integrations and these are performed
using Monte Carlo algorithms. The statistical errors are reasonable.

This article is organized as follows. After this introduction, we describe
briefly in the next section the $s=1/2$ antiferromagnetic Heisenberg quantum
spin chain in the framework of \cite{jimbo-miwa}, give the definition of the
dynamic structure factor $S$ and decompose it into $n$-spinon contributions $%
S_{n}$ with $n$ even. Then we write the expressions of $S_{2}$ and $S_{4}$
and give a brief account of their respective features and behaviors. In
section three we describe the five sum rules we use to estimate the
scattering weight of the four-spinon contribution and the results obtained.
As already mentioned, the calculations use Monte Carlo integration methods
and a discussion of the errors is given. Section four includes concluding
remarks and indicates few directions in which one can carry forward.

\section{Heisenberg chain and dynamic structure factor}

The antiferromagnetic $s=1/2$ XXX Heisenberg chain is defined as the
isotropic limit of the XXZ anisotropic Heisenberg model with the following
Hamiltonian: 
\begin{equation}
H=-\frac{1}{2}\hspace{-3pt}\sum_{n=-\infty }^{+\infty }\left( \sigma
_{n}^{x}\sigma _{n+1}^{x}\hspace{-3pt}+\sigma _{n}^{y}\sigma _{n+1}^{y}%
\hspace{-3pt}+\Delta \sigma _{n}^{z}\sigma _{n+1}^{z}\right) \,.
\label{hamiltonian}
\end{equation}%
$\Delta \equiv (q+q^{-1})/2$ is the anisotropy parameter and the isotropic
antiferromagnetic limit is obtained as $\Delta \rightarrow -1^{-}$, or
equivalently $q\rightarrow -1^{-}$. Here $\sigma _{n}^{x,y,z}$ are the usual
Pauli matrices acting at the site $n$ of the chain. Note that the definition
(\ref{hamiltonian}) and its isotropic limit are totally equivalent to the
usual one if we consider \cite{langari} the transformation $U=\exp \left(
i\pi J\sum_{j=-\infty }^{+\infty }s_{j}^{z}\right) $, which transforms the
Hamiltonian $H\left( J,\Delta ,h\right) =J\sum_{n=-\infty }^{+\infty }\left(
s_{n}^{x}\,s_{n+1}^{x}+s_{n}^{y}\,s_{n+1}^{y}+\Delta
\,s_{n}^{z}s_{n+1}^{z}+h\,s_{n}^{z}\right) $ into $H\left( J,-\Delta
,h\right) $, i.e., $UH\left( J,\Delta ,h\right) U^{-1}=-H\left( J,-\Delta
,h\right) $. As mentioned already in the introductory section, we take the
coupling constant $J=1$ as well as $\hbar =1$ and the external magnetic
field $h=0$. Lattice spacing is also taken equal to one. All this is
consistent with the notation and conventions used in \cite{jimbo-miwa} which
we follow closely.

The full exploitation of the quantum affine algebra $\mathrm{U}_{q}(\widehat{%
\mathrm{sl}}_{2})$ symmetry of the model requires an exact diagonalization
of the Hamiltonian directly in the thermodynamic limit. Because of two
different boundary conditions on the infinite chain, there are two
equivalent vacuum states $|0\rangle _{i}$, $i=0,1$. The Hilbert space $%
\mathcal{F}$ consists of $n$-spinon energy eigenstates $|\xi _{1},...,\xi
_{n}\rangle _{\epsilon _{1},...,\epsilon _{n};\,i}$ such that: 
\begin{equation}
H|\xi _{1},...,\xi _{n}\rangle _{\epsilon _{1},...,\epsilon
_{n};\,i}=\sum_{j=1}^{n}e(\xi _{j})|\xi _{1},...,\xi _{n}\rangle _{\epsilon
_{1},...,\epsilon _{n};\,i\;},  \label{energy-eigenstates}
\end{equation}%
where $e(\xi _{j})$ is the energy of spinon $j$ and $\xi _{j}$ is a spectral
parameter living on the unit circle. In the above relation, $\epsilon
_{j}=\pm 1$. The translation operator $T$ which shifts the spin chain by one
site acts on the energy eigenstates as follows: 
\begin{equation}
T|\xi _{1},...,\xi _{n}\rangle _{\epsilon _{1},...,\epsilon
_{n};\,i}=\prod_{i=1}^{n}\tau (\xi _{i})|\xi _{1},...,\xi _{n}\rangle
_{\epsilon _{1},...,\epsilon _{n};1-i\;},  \label{action-of-T}
\end{equation}%
where $\tau (\xi _{j})=e^{-ip(\xi _{j})}$ and $p(\xi _{j})$ is the lattice
momentum of spinon $j$. The exact expressions of $e(\xi _{j})$ and $p(\xi
_{j})$ are known \cite{faddeev-takhtajan,jimbo-miwa,ABS} and their isotropic
limits are given in eq (\ref{dispersion-relation}) below. The completeness
relation in $\mathcal{F}$ reads: 
\begin{equation}
\mathbf{I}=\sum_{i=0,1}\sum_{n\geq 0}\sum_{\{\epsilon _{j}=\pm 1\}_{j=1,n}}%
\frac{1}{n!}\oint \prod_{j=1}^{n}\frac{d\xi _{j}}{2\pi i\xi _{j}}\;|\xi
_{1},...,\xi _{n}\rangle _{\epsilon _{1},...,\epsilon _{n};\,i\;i;\,\epsilon
_{1},...,\epsilon _{n}}\langle \xi _{1},...,\xi _{n}|\,.
\label{completeness-relation}
\end{equation}

\subsection{The dynamic structure factor}

The dynamic structure factor we calculate is the zero-temperature limit of (%
\ref{definition DSF}) up to $2\pi $, namely the Fourier transform of the
transverse vacuum-to-vacuum two-point function defined by: 
\begin{equation}
S^{i,+-}(\omega ,k)=\int_{-\infty }^{\infty }dt\sum_{m\in \mathbb{Z}%
}e^{i(\omega t+km)}\,_{i}\langle 0|\sigma _{m}^{+}(t)\,\sigma
_{0}^{-}(0)|0\rangle _{i}\;,  \label{definition-of-S}
\end{equation}%
where $\omega $ is the neutron energy, always positive, and $k$ the neutron
momentum component along the chain. $\sigma ^{\pm }$ denotes $(\sigma
^{x}\pm i\sigma ^{y})/2$. The DSF satisfies the following relations: 
\begin{equation}
S(\omega ,k)=S(\omega ,-k)=S(\omega ,k+2\pi )\,,  \label{symmetries-of-DSF}
\end{equation}%
expressing reflection symmetry and periodicity. Inserting the completeness
relation (\ref{completeness-relation}) and using the Heisenberg relation: 
\begin{equation}
\sigma _{m}^{x,y,z}(t)=e^{iHt}\,T^{-m}\,\sigma
_{0}^{x,y,z}(0)\,T^{m}\,e^{-iHt}\,,  \label{heisenberg-relation}
\end{equation}%
we can write the transverse DSF (\ref{definition-of-S}) as the sum of $n$%
-spinon contributions, with $n$ even: 
\begin{equation}
S^{i,+-}(\omega ,k)=\sum_{n\,\,\mathrm{even}}S_{n}^{i,+-}(\omega ,k)\;,
\label{sum-over-n-spinons}
\end{equation}%
where the $n$-spinon DSF $S_{n}$ is given by: 
\begin{eqnarray}
S_{n}^{i,+-}(\omega ,k) &=&\frac{2\pi }{n!}\sum_{m\in \mathbb{Z}%
}\,\sum_{\epsilon _{1},...,\epsilon _{n}}\oint \prod_{j=1}^{n}\frac{d\xi _{j}%
}{2\pi i\xi _{j}}\,e^{im\left( k+\sum_{j=1}^{n}p_{j}\right) }\,\delta \left[
\omega -\sum\nolimits_{j=1}^{n}e_{j}\right] \,  \nonumber \\
&&\times \;X_{\epsilon _{n},...,\epsilon _{1}}^{i+m}(\xi _{n},...,\xi
_{1})\;X_{\epsilon _{1},...,\epsilon _{n}}^{1-i}(-q\xi _{1},...,-q\xi
_{n})\;,  \label{expression-of-Sn}
\end{eqnarray}%
a relation in which $X^{i}$ denotes the form factor: 
\begin{equation}
X_{\epsilon _{1},...,\epsilon _{n}}^{i}(\xi _{1},...,\xi _{n})\equiv
\,_{i}\langle 0|\sigma _{0}^{+}\left( 0\right) |\xi _{1},...,\xi _{n}\rangle
_{\epsilon _{1},...,\epsilon _{n};\,i}\;,  \label{form-factor}
\end{equation}%
and $i+m$ is to be read modulo 2. Note that each $S_{n}$ satisfies the
symmetry relations (\ref{symmetries-of-DSF}).

The form factor $X^{i}$ is known \cite{smirnov,jimbo-miwa}. It is expressed
as a trace of vertex operators in the context of the infinite-dimensional
representation of $\mathrm{U}_{q}(\widehat{\mathrm{sl}}_{2})$. The trace is
performed using $q$-deformed commutation relations of annihilation and
creation operators the vertex operators are expressed with. From the form
factor $X^{i}$ one can get a compact expression for the DSF $S_{n}$ \cite%
{ABS} in the general anisotropic case, an expression involving intricate
complex contour integrals. The isotropic limit, the one of interest in this
work, is obtained via the replacement \cite{jimbo-miwa,ABS}: 
\begin{equation}
\xi =ie^{-2i\varepsilon \rho }\,;\qquad q=-e^{-\varepsilon }\,,\qquad
\varepsilon \rightarrow 0^{+}\,,  \label{Isotropic-limit}
\end{equation}%
where $\rho $ is the new spectral parameter suited for this case. The
expressions of the energy $e$ and momentum $p$ in terms of $\rho $ are: 
\begin{equation}
e(\rho )=\frac{\pi }{\cosh (2\pi \rho )}=-\pi \sin \,p\;;\quad \cot
\,p=\sinh (2\pi \rho )\;;\quad -\pi \leq p\leq 0\;.
\label{dispersion-relation}
\end{equation}

\subsection{The two-spinon contribution}

The transverse two-spinon DSF $S_{2}$ is the less involved expression to
derive from (\ref{expression-of-Sn}). It has been obtained in \cite{BCK} and
reads: 
\begin{equation}
S_{2}(\omega ,k-\pi )=\frac{e^{-I(\rho )}}{4}\,\frac{\Theta (\omega -\omega
_{2l}\left( k\right) )\,\Theta (\omega _{2u}\left( k\right) -\omega )}{\sqrt{%
\omega _{2u}^{2}-\omega ^{2}}}\,\;.  \label{S2exact}
\end{equation}%
The notation of the dynamic structure factor has been eased since we will
deal only with the transverse DSF and the final results are independent of
the vacuum state chosen. The function $I(\rho )$ is given by: 
\begin{equation}
I(\rho )=\int_{0}^{+\infty }\frac{dt}{t}\frac{\cosh (2t)\,\cos (4\rho t)-1}{%
\,\sinh (2t)\cosh (t)}\,e^{t}\,,  \label{I-de-rho}
\end{equation}%
and $\omega _{2u}\left( k\right) $ and $\omega _{2l}\left( k\right) $ are
the familiar upper and lower\ bounds of the spin-wave continuum of
excitation energies given in (\ref{2 spinon continuum}). The spectral
parameter $\rho $ is related to $\omega $ and $k$ by the relation: 
\begin{equation}
\cosh \,\pi \rho =\sqrt{\frac{\omega _{2u}^{2}-\omega _{2l}^{2}}{\omega
^{2}-\omega _{2l}^{2}}}\,,  \label{relation-rho-omega-k}
\end{equation}%
which is obtained using eq (\ref{dispersion-relation}) and the
energy-momentum conservation laws: 
\begin{equation}
\omega =e_{1}+e_{2};\qquad k=-p_{1}-p_{2}.\text{\ \ }
\label{energy momentum conservation}
\end{equation}%
The properties of $S_{2}$ have been discussed in \cite{KMB,BKM} where a
thorough comparison with the M\"{u}ller ansatz (\ref{Muller ansatz}) is
carried. We will simply note that: (i) The confinement of $S_{2}$ in (\ref%
{S2exact}) to the spin-wave continuum means that the M\"{u}ller ansatz (\ref%
{Muller ansatz}) was in fact a description of the scattering weight from
two-spinon processes only, and the spin-wave continuum is actually a
`two-spinon continuum'. (ii) The exact two-spinon DSF $S_{2}$ exhibits a
manifest square-root singularity at the upper boundary of the two-spinon
continuum whereas the M\"{u}ller ansatz exhibits a square-root singularity
at the lower boundary.

\subsection{The four-spinon contribution}

The expression of the four-spinon DSF $S_{4}$ is given in \cite{ABS}. For $%
0\leq k\leq \pi $ it reads:%
\begin{equation}
S_{4}\left( \omega ,k-\pi \right) =C_{4}\int\limits_{-\pi
}^{0}dp_{3}\int\limits_{-\pi }^{0}dp_{4}\,F\left( \rho _{1},\rho _{2},\rho
_{3},\rho _{4}\right) .  \label{expression of S_4}
\end{equation}%
For other values of $k$, it extends by symmetry using (\ref%
{symmetries-of-DSF}). The notation is as follows. $C_{4}$ is a numerical
constant equal to:%
\begin{equation}
C_{4}=\frac{2^{-21}\cdot \pi ^{-14}}{3\left\vert \Gamma \left( \frac{1}{4}%
\right) \right\vert ^{8}\,\left\vert A\left( \frac{i\pi }{2}\right)
\right\vert ^{8}}=2.130627...\times \,10^{-7}\,,  \label{expression of C_4}
\end{equation}%
an expression in which $\Gamma $ is Euler's gamma function and:%
\begin{equation}
A\left( z\right) =\exp \left( -\int_{0}^{\infty }dt\frac{\sinh ^{2}t\left( 1-%
\frac{z}{i\pi }\right) \exp t}{t\sinh \left( 2t\right) \cosh t}\right) .
\label{function A}
\end{equation}%
The integrand $F$ in (\ref{expression of S_4}) has a rather complicated
expression. It writes: 
\begin{equation}
F\left( \rho _{1},\rho _{2},\rho _{3},\rho _{4}\right) =\sum\limits_{\left(
p_{1},p_{2}\right) }\frac{\exp \left[ -h\left( \rho _{1},\rho _{2},\rho
_{3},\rho _{4}\right) \right] \sum_{\ell =1}^{4}\left\vert g_{\ell }\left(
\rho _{1},\rho _{2},\rho _{3},\rho _{4}\right) \right\vert }{\sqrt{%
W_{u}^{2}-W^{2}}}.  \label{function F}
\end{equation}%
The different quantities involved in this expression are defined as follows: 
\[
W=\omega +\pi \left( \sin p_{3}+\sin p_{4}\right) ;\quad W_{u}=2\pi
\left\vert \sin K/2\right\vert ;\quad K=k+p_{3}+p_{4}; 
\]%
\begin{equation}
\cot p_{j}=\sinh \left( 2\pi \rho _{j}\right) ,\qquad -\pi \leq p_{j}\leq 0;
\label{ingredients in S_4}
\end{equation}%
the function $h$ in relation (\ref{function F}) reads: 
\begin{equation}
h\left( \rho _{1},\rho _{2},\rho _{3},\rho _{4}\right) =\sum\limits_{1\leq
i\leq j\leq 4}I\left( \rho _{ij}\right) ,  \label{function h}
\end{equation}%
where $\rho _{ij}=\rho _{i}-\rho _{j}$ and the function $I\left( \rho
\right) $ is given in (\ref{I-de-rho}). The function $g_{\ell }$ reads:%
\begin{eqnarray}
g_{\ell } &=&\left( -\right) ^{\ell +1}\left( 2\pi \right)
^{4}\sum\limits_{j=1}^{4}\cosh \left( 2\pi \rho _{j}\right)  \nonumber \\
&&\times \sum\limits_{m=\Theta \left( j-\ell \right) }^{4}\frac{%
\prod\limits_{i\neq \ell }\left( m-\frac{1}{2}\Theta \left( \ell -i\right)
+i\rho _{ji}\right) }{\prod\limits_{i\neq j}\pi ^{-1}\sinh \left( \pi \rho
_{ji}\right) }\prod\limits_{i=1}^{4}\frac{\Gamma \left( m-\frac{1}{2}+i\rho
_{ji}\right) }{\Gamma \left( m+1+i\rho _{ji}\right) }.  \label{function g_l}
\end{eqnarray}%
In relation (\ref{function F}), the sum is over the two pairs $\left(
p_{1},p_{2}\right) $ and $\left( p_{2},p_{1}\right) $ solutions of the
following energy-momentum conservation laws: 
\begin{equation}
W=-\pi \left( \sin p_{1}+\sin p_{2}\right) ;\qquad K=-p_{1}-p_{2}.
\label{conservation laws}
\end{equation}%
They read: 
\begin{equation}
\left( p_{1},p_{2}\right) =\left( -K/2+\arccos \left( W/\left[ 2\pi \sin
\left( K/2\right) \right] \right) ,-K/2-\arccos \left( W/\left[ 2\pi \sin
\left( K/2\right) \right] \right) \right) .
\label{solution of conservation laws}
\end{equation}%
$\allowbreak $Note that the solution in (\ref{solution of conservation laws}%
) is allowed as long as $W_{l}\leq W\leq W_{u}$ where $W_{u}$ is given in (%
\ref{ingredients in S_4}) and: 
\begin{equation}
W_{l}=\pi \left\vert \sin K\right\vert .  \label{W_l}
\end{equation}

The (analytic) behavior of the function $F$ in (\ref{function F}) was
investigated in \cite{ABS}. In particular, it was shown that the series $%
g_{\ell }$ is convergent for all values of its arguments and stays finite
when two spectral parameters or more get equal. Since the function $h$ goes
to $+\infty $ in these regions \cite{KMB}, the integrand $F$ of $S_{4}$ is
regular there. Furthermore, it was shown that $F$ is exponentially
convergent when one of the spectral parameters gets large, which means the
double integration over $p_{3}$ and $p_{4}$ in (\ref{expression of S_4})\ is
finite. All these analytic results pave the way for safe numerical
manipulations.

The behavior of $S_{4}$ as a function of the energy and momentum transfers $%
\omega $ and $k$ respectively is studied numerically in \cite{silakhal-abada}%
. First is determined the extent in the $\left( k,\omega \right) $-plane
outside which $S_{4}$ vanishes identically, i.e., the `four-spinon
continuum', by analogy with the two-spinon continuum. Then shapes of $S_{4}$
as a function of $\omega $ for different fixed values of $k$ and vice versa
are obtained. Consistency of $S_{4}$ in three areas is obtained: (i)\
Confinement to the independently determined four-spinon continuum. (ii)
Expected overall shape: sharp rise at the lower boundary of the continuum
followed by a longer tail at the upper boundary. (iii) Similarity with the
overall shape of the two-spinon contribution $S_{2}$.

\section{Sum rules for the dynamic structure function}

Once the expression of the four-spinon contribution to the total dynamic
structure factor is found and its behavior acquainted with, it is
interesting to have an estimate of its contribution. One way is to consider
sum rules, which are physical quantities related to the total dynamic
structure factor, generally via specific integrals, the values of which we
know exactly. This procedure is a reasonable good indicator of the weight of
each contribution since each $S_{n}$ is positive, and so is their total sum $%
S$.

For the present case, namely the one-dimensional nearest-neighbor Heisenberg
antiferromagnet, a number of these sum rules has been derived in \cite%
{hohenberg-brinkman}, see also \cite{muller,fledderjohann-karbach-mutter}.
The ones we consider in this work are: the static susceptibility, the
integrated intensity, the total integrated intensity, the first frequency
moment and the nearest-neighbor correlation function. In the sequel, for
each of these five sum rules, we calculate the corresponding contribution
from the two-spinon DSF $S_{2}$ and then the corresponding contribution from
the four-spinon DSF $S_{4}$. Then we make a weight comparison, with the
corresponding exact result when possible, or simply between the two results.

All the forthcoming calculations are numerical. The two-spinon calculations
are fairly straightforward and standard quadratures are sufficient. But
those related to $S_{4}$ are quite more involved and we use a Monte Carlo
algorithm for that. The $\chi ^{2}$ factor and the standard deviation $%
\sigma _{4}$ corresponding to each calculation are displayed. Finally, note
that all the exact results we refer to are known in the literature \cite%
{hohenberg-brinkman, muller}.

\subsection{Static susceptibility}

The first sum rule we consider is the one giving the \textit{static
susceptibility} $\varkappa \left( k\right) $ for two local spin operators.
It is related to the dynamic structure factor via the relation \cite{lovesey}%
:%
\begin{equation}
\varkappa \left( k\right) \equiv \frac{1}{\pi }\int\limits_{0}^{+\infty }%
\frac{d\omega }{\omega }S\left( \omega ,k\right) .  \label{chi-k}
\end{equation}%
The static susceptibility is known exactly in the limit $k\rightarrow 0$. It
is found to be: 
\begin{equation}
\varkappa \left( 0\right) \equiv \lim_{k\rightarrow 0}\varkappa \left(
k\right) =\frac{1}{2\pi ^{2}}.  \label{chi_0}
\end{equation}%
We therefore calculate the static susceptibility for both $S_{2}$ and $S_{4}$
using (\ref{chi-k}) for small values of $k$, which we denote $\varkappa
_{2}\left( k\right) $ and $\varkappa _{4}\left( k\right) $ respectively. Our
results are displayed in Table 1. Of course, $\varkappa _{2}\left( k\right) $
and $\varkappa _{4}\left( k\right) $ for the chosen values of small $k$ will
not compare directly to $\varkappa \left( 0\right) $. Note also that it is
not possible to take $k$ directly equal to zero before integration because
we would get zero identically. Finally, the behaviors of $\varkappa
_{2}\left( k\right) $ and $\varkappa _{4}\left( k\right) $ themselves\ as
functions of small $k$ are not of interest in this study.

One notes that the $\chi ^{2}$ factor stays around one but $\sigma
_{4}/\varkappa _{4}$ is relatively high with respect to the coming sum
rules, around 13\% on average. This can easily be traced to the sensitivity
of the integration to the low-energy region as there is the factor $1/\omega 
$. Indeed, better precision is obtained when $\omega $ appears in the
numerator and not in the denominator. For this sum rule, $\varkappa _{4}$ is
only compared to $\varkappa _{2}$ and on average, the ratio $\varkappa
_{4}/\varkappa _{2}$ is around 0.6\%. We note that the estimated
contribution of $S_{4}$ with respect to this sum rule is smaller than the
forthcoming ones.

\vbox{\begin{center}
\begin{tabular}{|c|c||c|c|c|c||c|c|}
\hline
$k$ & $\varkappa _{2}\left( k\right) $ & $\varkappa _{4}\left( k\right) $ & $\sigma _{4}$ & $\sigma _{4}/\varkappa _{4}$ & $\chi ^{2}$ & $\varkappa
\left( 0\right) $ & $\varkappa _{4}/\varkappa _{2}$ \\ \hline\hline
$0.1$ & 4.17340\ 10$^{-2}$ & 0.36927\ 10$^{-4}$ & 0.6344\ 10$^{-5}$ & 17.1\%
& 0.813 & 5.06606\ 10$^{-2}$ & 0.1\% \\ \hline
$0.2$ & 4.11826\ 10$^{-2}$ & 2.91744\ 10$^{-4}$ & 2.2114\ 10$^{-5}$ & 7.5\%
& 0.573 & 5.06606\ 10$^{-2}$ & 0.7\% \\ \hline
$0.3$ & 4.00307\ 10$^{-2}$ & 1.85373\ 10$^{-4}$ & 1.8943\ 10$^{-5}$ & 10.2\%
& 1.36 & 5.06606\ 10$^{-2}$ & 0.5\% \\ \hline
$0.4$ & 4.48920\ 10$^{-2}$ & 3.99597\ 10$^{-4}$ & 7.2076\ 10$^{-5}$ & 18.0\%
& 0.897 & 5.06606\ 10$^{-2}$ & 0.9\% \\ \hline
\end{tabular}
\end{center}

\begin{center}
Table 1: Static susceptibility for $S_{2}$ and $S_{4}$
\end{center}}

\subsection{Integrated intensity}

The next sum rule we consider is the one that defines the \textit{integrated
intensity} with respect to the neutron energy transfer, namely: 
\begin{equation}
I\left( k\right) \equiv \frac{1}{2\pi }\int\limits_{0}^{+\infty }d\omega
\,S\left( \omega ,k\right) .  \label{integrated intensity of k}
\end{equation}%
The behavior of the integrated intensity is not exactly known for all values
of $k$, but for small $k$, we know it is linear: 
\begin{equation}
I\left( k\right) \sim \frac{1.1\,k}{4\pi }\quad \left( k\;\mathrm{small}%
\right) .  \label{small k behavior of integrated intensity}
\end{equation}%
We will then use (\ref{integrated intensity of k}) to calculate, for small $%
k $, the integrated intensity $I_{2}\left( k\right) $ coming the two-spinon
DSF $S_{2}$ and the integrated intensity $I_{4}\left( k\right) $ coming from
the four-spinon DSF $S_{4}$. The results are displayed in Table 2 and
plotted in FIG 1.

We notice a clear linear behavior for $I_{2}\left( k\right) $, with a slope
roughly 75\% of that of $I\left( k\right) $. The four-spinon integrated
intensity presents a somewhat linear behavior too, with a slope roughly 1\%
of that of $I\left( k\right) $. The $\chi ^{2}$ factor is still of order one
and the standard deviation $\sigma _{4}$ is here too about 13\%, as for the
static susceptibility. The explanation lies also in the small $\omega $%
-region.

\vbox{\begin{center}
\begin{tabular}{|c|c||c|c|c|c||c|c|c|}
\hline
$k$ & $I_{2}(k)$ & $I_{4}(k)$ & $\sigma _{4}$ & $\sigma _{4}/I_{4}$ & $\chi
^{2}$ & $I(k)$ & $I_{2}/I$ & $I_{4}/I$ \\ \hline\hline
0.1 & 0.640435\ 10$^{-2}$ & 0.26462\ 10$^{-4}$ & 0.3656\ 10$^{-5}$ & 13.8\%
& 1.16 & 0.875535\ 10$^{-2}$ & 73.1\% & 0.3\% \\ \hline
0.2 & 1.27924\ 10$^{-2}$ & 1.72237\ 10$^{-4}$ & 1.7739\ 10$^{-5}$ & 10.2\% & 
1.30 & 1.750704\ 10$^{-2}$ & 80.2\% & 1.1\% \\ \hline
0.3 & 1.911885\ 10$^{-2}$ & 2.98517\ 10$^{-4}$ & 4.9236\ 10$^{-5}$ & 16.3\%
& 0.607 & 2.62606\ 10$^{-2}$ & 72.5\% & 1.1\% \\ \hline
0.4 & 2.643305\ 10$^{-2}$ & 3.11577\ 10$^{-4}$ & 4.0768\ 10$^{-5}$ & 13.1\%
& 0.951 & 3.501409\ 10$^{-2}$ & 75.0\% & 0.9\% \\ \hline
\end{tabular}
\end{center}

\begin{center}
Table 2: Integrated intensity for $S_{2}$ and $S_{4}$
\end{center}}

\FRAME{ftbphFUX}{3.378in}{4.4166in}{0pt}{\Qcb{Integrated intensity for $%
S_{2} $ and S$_{4}$ for small $k$.}}{\Qlb{Figure 1}}{i-k.eps}{\special%
{language "Scientific Word";type "GRAPHIC";maintain-aspect-ratio
TRUE;display "USEDEF";valid_file "F";width 3.378in;height 4.4166in;depth
0pt;original-width 8.1327in;original-height 10.6562in;cropleft "0";croptop
"1";cropright "1";cropbottom "0";filename '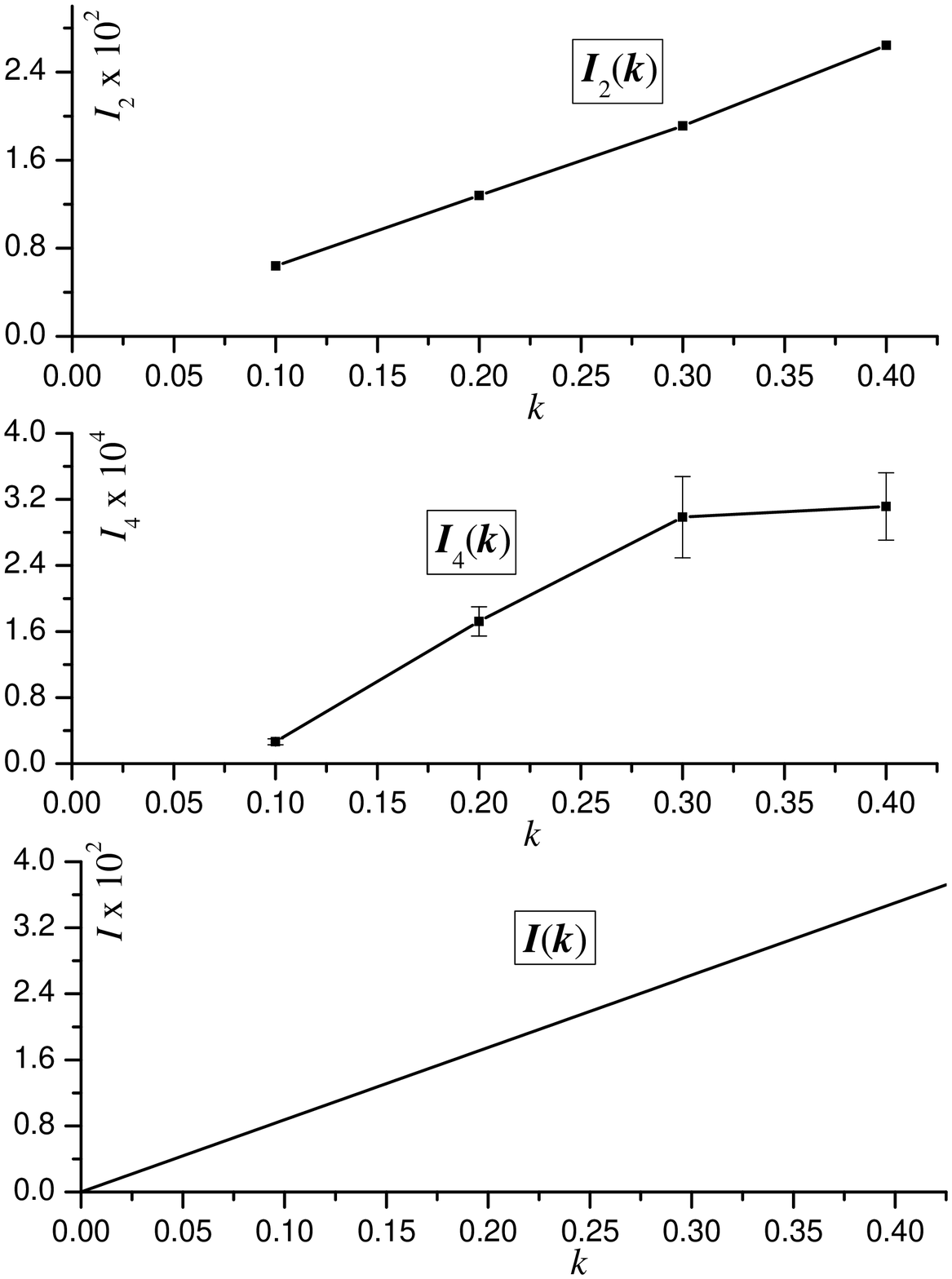';file-properties "XNPEU";}%
}

\subsection{Total integrated intensity}

The \textit{total integrated intensity} with respect to both energy and
momentum is defined by: 
\begin{equation}
I\equiv \frac{1}{\pi }\int\limits_{0}^{\pi }dkI\left( k\right) =\frac{1}{%
2\pi ^{2}}\int\limits_{0}^{\pi }dk\int\limits_{0}^{\infty }d\omega S\left(
\omega ,k\right) \text{.}  \label{integrated intensity}
\end{equation}%
It is known to be exactly equal to \cite{hohenberg-brinkman}: 
\begin{equation}
I=\frac{1}{4}.  \label{exact value of I}
\end{equation}%
Table 3 displays our results for this sum rule. We note a better standard
deviation: a relative value of 8\% only, compared to the previous 13\%. The $%
\chi ^{2}$ factor is of order one. The two-spinon contribution stays at
around 74\% whereas the four-spinon contribution is now at about 2.5\%. This
trend, namely a better precision and a 2.\% contribution, is going to
continue in the remaining sum rules.

\vbox{\begin{center}
\begin{tabular}{|c||c|c|c|c||c|c|c|}
\hline
$I_{2}$ & $I_{4}$ & $\sigma _{4}$ & $\sigma _{4}/I_{4}$ & $\chi ^{2}$ & $I$
& $I_{2}/I$ & $I_{4}/I$ \\ \hline\hline
0.1845412 & 0.00629916 & 0.0004973 & 7.9\% & 0.430 & 0.25 & 73.8\% & 2.5\%
\\ \hline
\end{tabular}
\end{center}

\begin{center}
Table 3: Total integrated intensity for $S_{2}$ and $S_{4}$
\end{center}}

\subsection{First frequency moment}

The next sum rule we look at is related to the \textit{first frequency moment%
}, defined by: 
\begin{equation}
K\left( k\right) \equiv \frac{1}{2\pi }\int\limits_{0}^{\infty }d\omega
\,\omega \,S\left( \omega ,k\right) .  \label{first-frequency-moment}
\end{equation}%
It is also known exactly, but for all $k$ this time. It reads: 
\begin{equation}
K\left( k\right) =\frac{8}{3}\left( \ln 2-\frac{1}{4}\right) \sin ^{2}\left( 
\frac{k}{2}\right) .  \label{exact K(k)}
\end{equation}%
We have carried out the runs for small values of the momentum transfer and
our results for the two-spinon contribution $K_{2}\left( k\right) $ and the
four-spinon contribution $K_{4}\left( k\right) $ are displayed in Table 4
below. They are also plotted in FIG 2.

On sees that $K_{2}\left( k\right) $ has the right quadratic behavior from (%
\ref{exact K(k)}) with a relative coefficient of roughly 69\%. But what is
remarkable is the fact that also $K_{4}\left( k\right) $ has the same
quadratic behavior, with a stable coefficient of about 2.5\%. The $\chi ^{2}$
factor is of order one and the standard deviation is better, around 6.7\%.
The better statistics comes partly from the fact that the neutron energy $%
\omega $ is present in the numerator and not in the denominator in the
integral (\ref{first-frequency-moment}), which makes the numerical
integration more stable close to $\omega =0$. One may then expect the higher
frequency moments $K_{i}\left( k\right) \equiv \frac{1}{2\pi }%
\int\limits_{0}^{\infty }d\omega \,\omega ^{i}\,S\left( \omega ,k\right) $
with $i>1$ to give even more stable results.

\vbox{\begin{center}
\begin{tabular}{|c|c||c|c|c|c||c|c|c|}
\hline
$k$ & $K_{2}\left( k\right) $ & $K_{4}\left( k\right) $ & $\sigma _{4}$ & $\sigma _{4}/K_{4}$ & $\chi ^{2}$ & $K\left( k\right) $ & $K_{2}/K$ & $K_{4}/K $ \\ \hline\hline
0.1 & 0.2053042\ 10$^{-2}$ & 0.80744\ 10$^{-4}$ & 0.6305\ 10$^{-5}$ & 7.8\%
& 1.196 & 0.2951853\ 10$^{-2}$ & 69.5\% & 2.7\% \\ \hline
0.2 & 0.813574\ 10$^{-2}$ & 3.08253\ 10$^{-4}$ & 2.1058\ 10$^{-5}$ & 6.8\% & 
1.549 & 1.177792\ 10$^{-2}$ & 69.1\% & 2.6\% \\ \hline
0.3 & 1.799619\ 10$^{-2}$ & 5.42950\ 10$^{-4}$ & 2.8636\ 10$^{-5}$ & 5.3\% & 
0.208 & 2.639001\ 10$^{-2}$ & 68.2\% & 2.1\% \\ \hline
0.4 & 3.294491\ 10$^{-2}$ & 12.70855\ 10$^{-4}$ & 8.8941\ 10$^{-5}$ & 7.0\%
& 0.524 & 4.664213\ 10$^{-2}$ & 70.6\% & 2.7\% \\ \hline
\end{tabular}
\end{center}

\begin{center}
Table 4: First frequency moment for $S_{2}$ and $S_{4}$
\end{center}}

\FRAME{ftbphFUX}{3.378in}{4.4166in}{0pt}{\Qcb{First frequency moment for $%
S_{2}$ and S$_{4}$ for small values of $k$.}}{\Qlb{Figure 2}}{k-k.eps}{%
\special{language "Scientific Word";type "GRAPHIC";maintain-aspect-ratio
TRUE;display "USEDEF";valid_file "F";width 3.378in;height 4.4166in;depth
0pt;original-width 8.1327in;original-height 10.6562in;cropleft "0";croptop
"1";cropright "1";cropbottom "0";filename '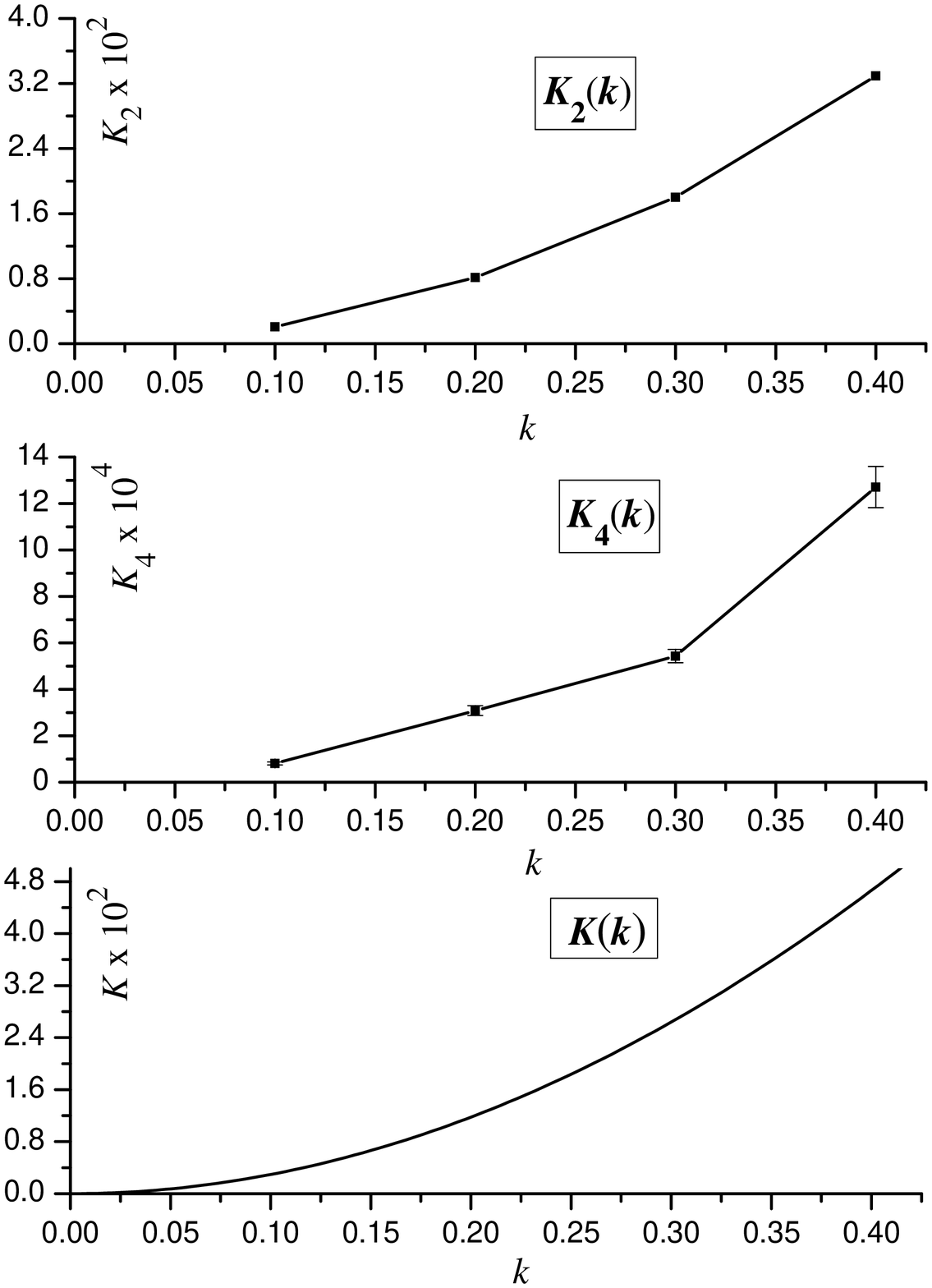';file-properties "XNPEU";}%
}

\subsection{Nearest-neighbor correlation}

The last sum rule we consider is related to the \textit{nearest-neighbor
correlation function} defined by the following relations: 
\begin{equation}
\left\langle \sigma _{n}^{z}\sigma _{n+1}^{z}\right\rangle \equiv \frac{1}{%
\pi }\int\limits_{0}^{\pi }dk\,\exp ^{-ik}\,I\left( k\right) =\frac{1}{2\pi
^{2}}\int\limits_{0}^{\pi }dk\exp ^{-ik}\int\limits_{0}^{\infty }d\omega
\,S\left( \omega ,k\right) .  \label{nearest neighbor correlation}
\end{equation}%
Its value is also known exactly \cite{hohenberg-brinkman}: 
\begin{equation}
\left\langle \sigma _{n}^{z}\sigma _{n+1}^{z}\right\rangle =\frac{E_{G}}{3},
\label{exact value of nn correlation}
\end{equation}%
where $E_{G}$ is the ground-state energy per site which, in the
thermodynamic limit, is given by: 
\begin{equation}
E_{G}=-\left( \ln 2-\frac{1}{4}\right) .  \label{ground state energy}
\end{equation}%
Our results for the nearest-neighbor correlation function are displayed in
Table 5 below. The two-spinon contribution is about 72\% \ and the
four-spinon contribution is around 2.4\%. The (relative) standard deviation $%
\sigma _{4}$ is about 12\% and the $\chi ^{2}$ factor is of order one.

\vbox{\begin{center}
\begin{tabular}{|c||c|c|c|c||c|c|c|}
\hline
$\left\langle \sigma _{n}^{z}\sigma _{n+1}^{z}\right\rangle _{2}$ & $\left\langle \sigma _{n}^{z}\sigma _{n+1}^{z}\right\rangle _{4}$ & $\sigma
_{4}$ & $\sigma _{4}/\left\vert \left\langle {}\right\rangle _{4}\right\vert 
$ & $\chi ^{2}$ & $\left\langle \sigma _{n}^{z}\sigma
_{n+1}^{z}\right\rangle $ & $\left\langle {}\right\rangle _{2}/\left\langle
{}\right\rangle $ & $\left\langle {}\right\rangle _{4}/\left\langle
{}\right\rangle $ \\ \hline\hline
$-0.105594$ & $-0.0035309$ & 0.000421 & 11.9\% & 0.580 & $-0.1477157$ & 
71.5\% & 2.4\% \\ \hline
\end{tabular}
\end{center}

\begin{center}
Table 5: Nearest-neighbor correlation for $S_{2}$ and $S_{4}$
\end{center}}

\section{Conclusion and outlook}

In this work, we have calculated five sum rules for the exact four-spinon
dynamic structure factor $S_{4}$ in the spin 1/2 Heisenberg
antiferromagnetic quantum spin chain. The calculations were all numerical
using Monte Carlo integration methods. The statistics is satisfactory: the $%
\chi ^{2}$ factor is always around one and the standard deviation $\sigma
_{4}$ range between 7\% and 13\% of the calculated quantity. It is better
for sum rules in which the neutron momentum transfer $\omega $ appears in
the numerator since this make the integration more stable when close to zero.

From this analysis, we conclude that the four-spinon DSF $S_{4}$ contributes
to the total dynamic structure factor $S$ a factor between 1\% and 2.5\%,
depending on the sum rule used, whereas the two-spinon contribution is
between 70\% and 75\%. These results are consistent with the expectation
from finite chain calculations \cite{muller-thomas-beck-bonner} that the
spectral weight of states outside the two-spinon continuum is roughly two
orders of magnitude smaller than the spectral weight of neighboring states
within the continuum.

There are five directions in which one may wish to carry forward. The first
direction is to try to determine an expression for the six-spinon
contribution $S_{6}$, study its behavior as a function of $\omega $ and $k$
and then determine its contribution to the total $S.$ But one should know
that, technically, matters may be more involved.

The second direction to explore is the study of the dynamic structure factor
in the anisotropic case. This is also of physical interest since perfect
isotropy is only an ideal limit. The model is exactly solvable and we do
have generic expressions for $S_{n}$ in the form of contour integrals in the
spectral parameters' complex planes \cite{bougourzi2}. The difficulty here
is that the integrands involve much more complicated functions which are
already present in $S_{2}$, and one should expect intricate complexities in
this more general case.

The third direction to explore is to include a (small) external magnetic
field. There are finite-chain calculations in this regard, \cite%
{muller-thomas-beck-bonner} and more recently \cite{KBM,KM}. But one has to
remember that the model in not exactly solvable in this case. One will then
have to try small perturbations around the zero-field limit solution.

The fourth direction is the finite-temperature case. Here too there are
finite-chain \cite{muller-thomas-beck-bonner} and field-theory results \cite%
{schulz}. It is then certainly interesting to see the temperature effects on 
$S_{2}$ and perhaps on $S_{4}$.

The fifth direction is to look into the situation of a spin-one chain. The
model is still exactly solvable and, exploiting the quantum group symmetry,
compact expressions for the form factors are available \cite%
{idzumi,bougourzi-weston,bougourzi-any-spin}. One key issue in this regard
is to try to recover the Haldane gap \cite{haldane} through these exact
manipulations.

\begin{acknowledgments}
A substantial part of the numerical work was done at the Abdu-Salam ICTP,
Trieste, through the Junior Associateship program. B.S. warmly thanks the
Center for this.
\end{acknowledgments}


\begin{thebibliography}{99}
\bibitem{heisenberg} W. Heisenberg, Z. Phys. 49 (1928) 619.

\bibitem{hirakawa-kurogi} K. Hirakawa and Y. Kurogi, Prog. Theor. Phys. 
\textbf{46} (1970) 147.

\bibitem{okazaki1} A. Okazaki, J. Phys. Soc. Japan \textbf{26} (1969) 870.

\bibitem{okazaki2} A. Okazaki, J. Phys. Soc. Japan \textbf{27} (1969) 518.

\bibitem{baxter} R.J. Baxter, `\textit{Exactly Solved Models in Statistical
Mechanics}', Academic Press, 1982.

\bibitem{anderson1} P.W.\ Anderson, Phys. Rev. 86 (1952) 694.

\bibitem{bethe} H.A. Bethe, Z. Phys. 71 (1931) 205.

\bibitem{des cloizeaux-pearson} J. des Cloizeaux and J.J. Pearson, Phys.
Rev. \textbf{128} (1962) 2131.

\bibitem{endoh-shirane-birgeneau-richards-holt} Y. Endoh, G. Shirane, R.J.
Birgeneau, P.M. Richards and S.L. Holt, Phys. Rev. Lett. \textbf{32} (1974)
170.

\bibitem{heilmann-shirane-endoh-birgeneau-holt} I.U. Heilmann, G. Shirane,
Y. Endoh, R.J. Birgeneau and S.L.\ Holt, Phys. Rev. \textbf{B}18 (1978) 3530.

\bibitem{hutchings-ikeda-milne} M.T. Hutchings, H. Ikeda and J.M. Milne, J.
Phys. C 12 (1979) L739.

\bibitem{satija-axe-shirane-yoshizawa-hirakawa} S.K. Satija, J.D. Axe, G.
Shirane, H. Yoshizawa and K. Hirakawa, Phys. Rev. \textbf{B}21 (1980) 2001.

\bibitem{faddeev-takhtajan} L.D. Faddeev and L.A. Takhtajan, Phys. Lett. 
\textbf{A}85 (1981) 375.

\bibitem{fowler} M. Fowler, Phys. Rev. \textbf{B}18 (1978) 421.

\bibitem{anderson2} P.W.\ Anderson, Science 235 (1987) 1196.

\bibitem{haldane} F.D.M. Haldane, Phys.\ Rev.\ Lett. \textbf{50} (1983) 1153.

\bibitem{yamada} T. Yamada, Prog. Theor. Phys. \textbf{41} (1969) 880.

\bibitem{bonner-sutherland-richards} J.C. Bonner, B. Sutherland and P.M.
Richards, in Proceeding of the 20th Annual Conference on Magnetism and
Magnetic Materials, Edited by C.D. Graham, G.H. Lander and J.J. Rhyne, AIP,
New York, 1975.

\bibitem{lovesey} S.W. Lovesey, `\textit{Theory of neutron scattering from
condensed matter'}, Clarendon, Oxford, 1987.

\bibitem{squires} G.L. Squires, `\textit{Introduction to the theory of
thermal neutron scattering'}, Cambridge University Press, 1996.

\bibitem{nagler-tennant-cowley-perring-satija} S.E. Nagler, D.A.\ Tennant,
R.A. Cowley, T.G. Perring and S.K. Satija, Phys. Rev. \textbf{B}44 (1991)
12361.

\bibitem{kretzen-mikeska-patzak} H.H. Kretzen, H.J. Mikeska and E. Patzak,
Z. Phys. \textbf{271} (1974) 269.

\bibitem{holstein-primakoff} T. Holstein and H. Primakoff, Phys. Rev. 
\textbf{58} (1940) 1098.

\bibitem{tennant-nagler-welz-shirane-yamada} D.A. Tennant, S.E. Nagler, D.
Weltz, G. Shirane and K. Yamada, Phys. Rev. \textbf{B}52 (1995) 13381.

\bibitem{schulz} H.J. Schulz, Phys. Rev. \textbf{B}34 (1986) 6372.

\bibitem{tennant-perring-cowley-nagler} D.A. Tennant, T.G. Perring, R.A.
Cowley and S.E.\ Nagler, Phys. Rev. Lett. \textbf{70} (1993) 4003.

\bibitem{tennant-cowley-nagler-tsvelik} D.A. Tennant, R.A. Cowley, S.E.
Nagler and A.M. Tsvelik, Phys. Rev. \textbf{B}52 (1995) 13368.

\bibitem{muller-thomas-beck-bonner} G. M\"{u}ller, H. Thomas, H. Beck and
J.C. Bonner, Phys. Rev. \textbf{B}24 (1981) 1429.

\bibitem{deisz-jarrell-cox} J. Deisz, M. Jarrell and D.L. Cox, Phys. Rev. 
\textbf{B}42 (1990) 4869.

\bibitem{viswanath} V.S. Viswanath \textit{et al}., Phys. Rev. \textbf{B}49
(1994) 9702.

\bibitem{luther-peschel} A. Luther and I. Peschel, Phys. Rev. \textbf{B}9
(1974) 2911.

\bibitem{korepin-izergin-bogoliubov} V.E. Korepin, A.G.\ Izergin and N.M.
Bogoliubov, `\textit{The Quantum Inverse Scattering Method and Correlation
Functions}', Cambridge University Press, 1993.

\bibitem{difrancesco-mathieu-senechal} P. Di Francesco, P. Mathieu and D. S%
\'{e}n\'{e}chal, `\textit{Conformal Field Theory}', Springer, 1997.

\bibitem{frenkel-jing} I.B. Frenkel and N.H.\ Jing, Proc. Natl. Acad. Sci.
85 (1988) 9373.

\bibitem{davies-foda-jimbo-miwa-nakayashiki} O. Davies, O. Foda, M. Jimbo,
T. Miwa and A. Nakayashiki, Comm. Math. Phys. 151 (1993) 89.

\bibitem{abada-bougourzi-gradechi} A. Abada, A.H.\ Bougourzi and M.A. El
Gradechi, Mod. Phys. Lett. A8 (1993)\ 715.

\bibitem{bougourzi1} A.H. Bougourzi, Nucl. Phys. \textbf{B}404 (1993)\ 457.

\bibitem{jimbo-miwa} M. Jimbo and T. Miwa, `\textit{Algebraic Analysis of
Solvable Lattice Models}', American Mathematical Society, 1994.

\bibitem{bougourzi2} A.H. Bougourzi, Mod. Phys. Lett \textbf{B}10 (1996)
1237.

\bibitem{BCK} A.H. Bougourzi, M. Couture and M. Kacir, Phys. Rev. \textbf{B}%
54 (1996) 12669.

\bibitem{KMB} M. Karbach, G. M\"{u}ller and A.H. Bougourzi, `\textit{%
Two-spinon dynamic structure factor of the one-dimensional $S=1/2$
Heisenberg antiferromagnet}', \texttt{cond-mat/9606068}.

\bibitem{BKM} A.H. Bougourzi, M. Karbach and G. M\"{u}ller, `\textit{Exact
two-spinon dynamic structure factor of the one-dimensional} $s=1/2$ \textit{%
Heisenberg-Ising antiferromagnet}', \texttt{cond-mat/9712101}.

\bibitem{ABS} A. Abada, A.H. Bougourzi and B. Si-lakhal, Nucl. Phys. \textbf{%
B}497 [FS] (1997) 733.

\bibitem{silakhal-abada} B. Si Lakhal and A. Abada, J. Phys. A: Math. Gen. 
\textbf{37} (2004) 497.

\bibitem{langari} A. Langari, Phys. Rev. \textbf{B}58 (1998) 14467.

\bibitem{smirnov} F.A. Smirnov, `\textit{Form Factors in Completely
Integrable Models of Quantum Field Theory}' World Scientific, Singapore,
1992.

\bibitem{hohenberg-brinkman} P.C. Hohenberg and W.F. Brinkman, Phys. Rev. 
\textbf{B}10 (1974) 128.

\bibitem{muller} G. M\"{u}ller, Phys. Rev. \textbf{B}26 (1982) 1311.

\bibitem{fledderjohann-karbach-mutter} A. Fledderjohann, M. Karbach and
K.-H. M\"{u}tter, Phys. Rev. \textbf{B}53 (1996) 11543.

\bibitem{KBM} M. Karbach, D. Biegel and G. M\"{u}ller, `\textit{%
Quasiparticles governing the zero-temperature dynamics of the 1D spin-1/2
Heisenberg antiferromagnet in a magnetic field}', \texttt{cond-mat/0205142}.

\bibitem{KM} M.\ Karbach and G. M\"{u}ller, `\textit{Line shape predictions
via Bethe ansatz for the one dimensional spin-1/2 Heisenberg antiferromagnet
in a magnetic field}', \texttt{cond-mat/0005174}.

\bibitem{idzumi} M. Idzumi, Int. J. Mod. Phys. \textbf{A}9 (1994) 4449; `%
\textit{Correlation functions of the spin 1 analog of the XXZ model}', 
\texttt{hep-th/9307129}.

\bibitem{bougourzi-weston} A.H. Bougourzi and R.A. Weston, Nucl. Phys. 
\textbf{B}417 (1994) 439.

\bibitem{bougourzi-any-spin} A.H. Bougourzi, `\textit{Bosonization of
quantum affine groups and its application to higher spin Heisenberg model}', 
\texttt{q-alg/9706015}.
\end{thebibliography}
\end{document}